\documentclass[graybox]{svmult}

\usepackage{mathptmx}       
\usepackage{helvet}         
\usepackage{courier}        
\usepackage{type1cm}        
\usepackage{makeidx}         
\usepackage{graphicx}        
\usepackage{multicol}        
\usepackage[bottom]{footmisc}

\begin{document}

\title*{Virtual Observatory activities in the AMIGA group}
\titlerunning{VO activities in the AMIGA group}
\author{Ru\'iz, J.E., Santander-Vela, J.D., Garc\'ia, E., Espigares, V., Leon, S., Verdes-Montenegro, L.}
\authorrunning{Ru\'iz, J.E. et al.}
\institute{Jos\'e Enrique Ru\'iz \at IAA-CSIC, Camino Bajo de Hu\'etor 50, 18008 Granada Spain, \email{jer@iaa.es}
\and Juan de Dios Santander Vela \at IAA-CSIC, Camino Bajo de Hu\'etor 50, 18008 Granada Spain, \email{jdsant@iaa.es}
\and Emilio Garc\'ia \at IAA-CSIC, Camino Bajo de Hu\'etor 50, 18008 Granada Spain, \email{garcia@iaa.es}
\and Victor Espigares \at IAA-CSIC, Camino Bajo de Hu\'etor 50, 18008 Granada Spain, \email{espigar@iaa.es}
\and St\'ephane Leon \at IRAM, Avenida Divina Pastora, 7, 18012 Granada Spain, \email{leon@iram.es}
\and Lourdes Verdes-Montenegro \at IAA-CSIC, Camino Bajo de Hu\'etor 50, 18008 Granada Spain, \email{lourdes@iaa.es}}
%
%
\maketitle

\abstract*{The AMIGA project (Analysis of the interstellar Medium of Isolated GAlaxies) is an international scientific collaboration led from Instituto de Astrofisica de Andalucia (CSIC). The group's experience in radio astronomy databases turned, as a natural evolution, into an active participation in the development of data archives and radio astronomy software. The contributions of the group to the Virtual Observatory have been mostly oriented towards the deployment of large VO compliant databases and the development of access interfaces (IRAM 30m Pico Veleta, DSS-63 70m in Robledo de Chavela). We also have been working in the development of an API for VO tools that will ease access to VO registries and communication between different VO software. A collaboration with the Kapteyn Astronomical Institute has started recently in order to perform a complete renovation of the only existing high-level software (GIPSY) for the analysis of datacubes, allowing its fully integration in the VO.}

\abstract{The AMIGA project (Analysis of the interstellar Medium of Isolated GAlaxies) is an international collaboration led from the Instituto de Astrof\'isica de Andaluc\'ia (CSIC). The group's experience in radio astronomy databases turned, as a natural evolution, into an active participation in the development of data archives and radio astronomy software. The contributions of the group to the VO have been mostly oriented towards the deployment of large VO compliant databases and the development of access interfaces (IRAM 30m Pico Veleta, DSS-63 70m in Robledo de Chavela). 
\\\\
We also have been working in the development of an API for VO tools that will ease access to VO registries and communication between different VO software. A collaboration with the Kapteyn Astronomical Institute has started recently in order to perform a complete renovation of the only existing high-level software (GIPSY) for the analysis of datacubes, allowing its fully integration in the VO.}

\section{Introduction}
\label{sec:1}
The AMIGA\footnote{Analysis of the interstellar Medium of Isolated GAlaxies} project is an international scientific collaboration led from the Instituto de Astrof\'isica de Andaluc\'ia (CSIC). It focuses on a multiwavelength analysis of the interstellar medium of an statistically significant sample of isolated galaxies, in order to provide a pattern of behaviour to the study of galaxies in denser environments. 

The compilation of a strictly selected sample of more than 1000 isolated galaxies, including optical, IR, radio line and continuum measures has formed a group with an important experience in access to radio astronomical databases. At the same time, the exploitation of the data justifies the need to develop high-level tools for accessing and analyzing data coming from the Virtual Observatory. 

\section{The AMIGA VO Catalog}
\label{sec:2}
The AMIGA catalog is fully accessible from a web interface\footnote{http://www.iaa.es/AMIGA.html} or via ConeSearch~\cite{conesearch} VO service registered in ESA-VO Regsitry\footnote{http://esavo.esa.int/registry/}. In the web access interface the data can be requested either searching by the object's name or using a combination of parameters (distance, velocity, etc.) in order to produce your own subsample, in HTML, ASCII or VOTable~\cite{votable} format. The interface provides some links that allow the output to be sent to Topcat~\cite{topcat} or Aladin~\cite{aladin}. The AMIGA VO Catalog has been registered in the DCA Census of European Data Centres\footnote{http://cds.u-strasbg.fr/twikiDCA/bin/view/EuroVODCA/DCACensus}.

\begin{figure}
\sidecaption
\includegraphics{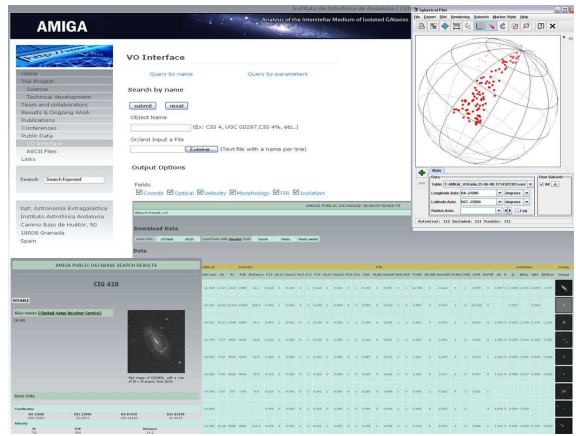}
\caption{The AMIGA VO Catalog web interface. User-friendly query forms allow access to a large variety of data where output format and the type of data requested may be chosen by the user. The output can be sent to other VO software, not necessarily locally installed, in order to analyze the information requested. The AMIGA VO Catalog is frequently updated with new data.}
\label{fig:1}
\end{figure}

\section{Archives}
\label{sec:3}
The AMIGA group expertise in access to radio astronomical databases and the implementation of the AMIGA VO Catalog described in Sect.~\ref{sec:2} led, as a next step, to the deployment of VO compliant archives and the development of access interfaces and VO services. 

\subsection{RADAMS}
\label{subsec:3.1}
As no suitable data model existed for single-dish radio astronomy, and in order to develop a VO-compliant archive for single-dish antennas, a complete radio data model had to be defined. RADAMS~\cite{radams} (Radio Astronomical DAta Model for Single-dish multiple-feed telescopes) is the first proposed VO data model for single-dish observations. It has been released as an IVOA note\footnote{http://www.ivoa.net/cgi-bin/twiki/bin/view/IVOA/RADAMS} within the Current Data Modelling Efforts.

It is based upon existing IVOA\footnote{International Virtual Observatory Alliance} data models, but it specifies the linking between those different data models, plus all the attributes that are needed for proper archival and retrieval of single dish observations, such as observing mode, switching mode, and the like. Its design comes from the development of a data model for the DSS-63 antenna archive, plus feedback from our team mates from IRAM to better adapt it to the IRAM 30m antenna at Pico Veleta.

\subsection{The DSS-63 VO Archive}
\label{subsec:3.2}
The development of the DSS-63 VO archive is being done in collaboration with the LAEFF-INTA. The station DSS-63 is the largest antenna in MDSCC\footnote{Madrid Deep Space Communication Complex} in Robledo de Chavela. According to an international agreement, up to 5\% of its operational time is routinely scheduled for radio astronomical observations in K-band (18 to 26 GHz).

Every year, raw and reduced data associated to observations made by DSS-63 70m antenna are saved in logfiles and FITS~\cite{fits} files. The DSS-63 VO Archive is filled thanks to a \texttt{datafiller} developed in Python scripting which also converts DSS-63 FITS files to standard formatted spectral FITS files readable by VO tools as VOSpec~\cite{vospec}. The data will be accessed from a web interface and through ConeSearch~\cite{conesearch} and SSAP~\cite{ssap} VO services.

\subsection{The IRAM 30m VO Archive}
\label{subsec:3.3}
Observations associated to more than 200 scientific projects are carried out each year in the IRAM 30m antenna. This generates more than 1 TB of data covering three spectral bands in radio frequencies 80-115 GHz, 130-183 GHz and 200-280 GHz where mostly molecular rotational transitions can be detected. 

The development of the IRAM 30m VO archive is being done in collaboration with IRAM. TAPAS\footnote{Telescope Archive Public Access System} is an archive of headers data, the data model is based on RADAMS~\cite{radams} and on IRAM 30m NCS\footnote{New Control System} data structure. The archive is filled in real time thanks to a \texttt{datafiller} developed in Python scripting. A prototype for the web access interface has been designed following scientific use cases. It will allow data retrieval via user-friendly forms, in addition to ConeSearch~\cite{conesearch} and SSAP~\cite{ssap} VO services.

\begin{figure}
\sidecaption
\includegraphics{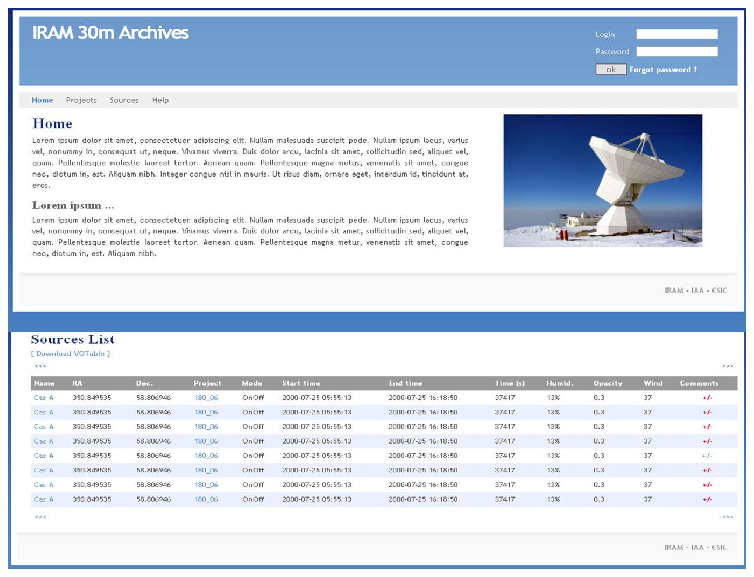}
\caption{Based upon use cases requested among advanced users, the IRAM 30m VO archive interface prototype is the template for the upcoming functional web interface. It will allow accessing headers data from projects, sources, scans, etc. The output formats will be chosen by the user (VOTable, HTML, ASCII) besides the filter criteria. Developed following VO standards it will bring in addition VO services for external querying.}
\label{fig:2}
\end{figure}

\section{Tools}
\label{sec:4}
Interoperability~\cite{interoperability} is the \emph{Rosetta Stone} of the Virtual Observatory. It does not only allow concurrent access to distributed, heterogeneous data, but also enables VO soft to communicate. The  SAMP~\cite{samp} (Simple Access Message Protocol) offers new born VO tools with all the functionalities coming from existing VO packages. Since they can communicate and share data sets, they form a huge VO meta-software in a continuously evolving ecosystem. 

\subsection{MOVOIR}
\label{subsec:4.1}
We are working on the development and application of techniques that will ease interoperability and communication in the VO context. The MOVOIR\footnote{MOdular Virtual Observatory Interface for Radio-astronomy} is a ongoing effort that aims to combine existing VO open source tools (Astro Runtime~\cite{astroruntime}, Plastic~\cite{plastic} client and server, and STIL~\cite{topcat} library) in order to produce an easily embeddable modular interface for radio astronomy tools and providing clean interfacing with the RADAMS~\cite{radams}.

MOVOIR is being implemented in different software packages like MASSA\footnote{MAdrid Single Spectra Analysis} and MADCUBA\footnote{MAdrid Data CUBe Analysis} from the DAMIR\footnote{Department of Molecular and Infrared Astrophysics - CSIC} group. 

\begin{figure}
\includegraphics[scale=0.9]{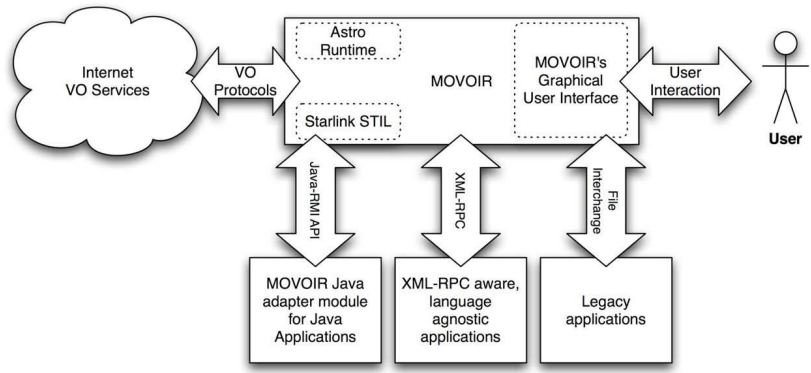}
\caption{MOVOIR's architecture.}
\label{fig:3}
\end{figure}

\subsection{GIPSY: a multidimensional future}
\label{subsec:4.2}
The scientific goals of the AMIGA project are based on the analysis of a significant amount of datacubes. We have started a collaboration with the Kapteyn Astronomical Institute in order to develop a new VO compliant package, including present GIPSY's~\cite{gipsy} core functionalities and new ones based on use cases elaborated with advanced users. One of the main goals is to provide local interoperability between GIPSY and other VO tools. In addition, the connectivity with the VO environment will provide access to 3D data VO archives, and ALMA datacubes in particular. For this purpose, a special effort need to be done in order to provide mature standards for datacubes access VO protocols.

%
%
%

\end{document}